\documentclass[aps,prb,twocolumn,superscriptaddress,showpacs,floatfix]{revtex4-1}

\usepackage{graphicx,color}
\usepackage{amsfonts}
\usepackage[figuresright]{rotating}  
\usepackage{amssymb}
\usepackage{amsmath}
\usepackage{psfrag}
\usepackage{subfigure}
\usepackage{multirow}
\usepackage{tabularx}
\usepackage{textcomp}
\usepackage{units}
\usepackage{hyperref}

\def\nn{\nonumber}

\def\beq{\begin{eqnarray}}
\def\eeq{\end{eqnarray}}

\renewcommand{\v}[1]{\ensuremath{\mathbf{#1}}} 
 
\let\baraccent=\= 
\renewcommand{\=}[1]{\stackrel{#1}{=}} 

\begin{document}
\title{Driven Electronic States at the Surface of  a Topological Insulator}
\author{Benjamin M.\ \surname{Fregoso}}
\affiliation{Department of Physics, University of California, Berkeley, Berkeley CA, 94720, USA} 
\affiliation{Joint Quantum Institute and Condensed Matter Theory Center, Department of Physics, University of Maryland, College Park, Maryland 20742-4111, USA} 

\author{Y.H. \ \surname{Wang}}
\affiliation{Department of Physics, Massachusetts Institute of Technology, Cambridge, Massachusetts 02139, USA} 
\affiliation{Department of Applied Physics, Stanford University, Stanford, CA 94305, USA} 

\author{N. \ \surname{Gedik}}
\affiliation{Department of Physics, Massachusetts Institute of Technology, Cambridge, Massachusetts 02139, USA} 

\author{Victor \ \surname{Galitski}}
\affiliation{Joint Quantum Institute and Condensed Matter Theory
  Center, Department of Physics, University of Maryland, College Park,
  Maryland 20742-4111, USA} 

\begin{abstract}
Motivated by recent photoemission experiments on the surface of topological insulators we compute the spectrum of driven topological surface excitations in the presence of an external light source. 
We completely characterize the spectral function of these non-equilibrium electron excitations for both linear and circular polarizations of the incident light. 
We find that in the latter case, the circularly polarized light gaps out the surface states, 
whereas linear polarization gives rise to an anisotropic metal with multiple Dirac cones.
We compare the sizes of the gaps with recent pump-probe photoemission measurements and find good agreement.
We also identify theoretically several new features in the time-dependent spectral function, such as shadow Dirac cones.
\end{abstract}

\maketitle

\section{Introduction}
Topological properties of matter usually manifest in the appearance of electronic states at the boundary\cite{Tsui1982, *Volovik1992,*Su1979}.
An especially interesting class of such topological boundary modes arises in three dimensional 
(3D) topological insulators (TI) which have now been detected experimentally in several 
material systems\cite{Hasan2010,Roy2009,Moore2007,Fu2007, Dzero2010,*Wolgast,*Zhang2013}.
The edge states consist of fermions in 2D with linear dispersion relation and where its spin and momenta have a 
fixed relative orientation.  
Recently, a new possibility for creating topological band structures in non-equilibrium was suggested~\cite{Lindner2011,Kitagawa2010,Rudner2013}, 
where an initially topologically trivial semiconductor is converted into a topological insulator via an external irradiation. The resulting state was dubbed
a Floquet topological insulator and an analogue to such a state was recently  realized experimentally in a photonic system\cite{Rechtsman2013}.
In the same vein, other theoretical works\cite{Kitagawa2011,Oka2009,Gu2011,*Dora2012} have studied the realization of a lattice quantum Hall state with time periodic
perturbations.

The focus of these previous studies has been on turning an electronic system with a topologically-trivial band structure into a topological insulator
by subjecting it to a periodic-in-time perturbation. Here, on the contrary, we study the effect of an external irradiation on an initially topological state. 
The motivation comes from the development of new experimental probes that make it possible to access the time-resolved excitation spectrum 
of driven electrons at the surface of TIs using time-resolved photoemission spectroscopy. \cite{Wang2012}
Below, we focus specifically on  the properties of driven Dirac electrons on the surface of existing 3D topological materials 
such as Bi$_x$Sb$_{1-x}$ alloy, Bi$_2$Te$_3$ and  Bi$_2$Se$_3$ (Ref.~\onlinecite{Hsieh2008,*Xia2009,*Zhang2009}). 
We are particularly interested in describing the modification of the spectrum of the boundary modes due to the irradiation as a function of the parameters of the incident light. 
The spectrum is composed of Floquet bands of the driven Dirac Hamiltonian, as discussed  in the previous related works \onlinecite{Syzranov2008,Oka2009,Zhou2011,Gomez-Leon2013}.
Besides, proving the existence of new dynamical Dirac cones which can be engineered, 
we also unify and extend previous analysis and apply our results specifically to describe an experiment, which observed an induced energy gap in driven
surface states of Bi$_2$Se$_3$ using time- and- angle-resolved photoemission spectroscopy (TrARPES) on  (Ref.~\onlinecite{Wang}). 
We find that our results fit the data well. 

\section{Model of driven surface states} 
\label{sec:model_driven_ti}
We consider non-interacting electrons at   
the surface of a TI with incident light normal to the surface. The Hamiltonian is,
\beq
H(\v{k},t) &=& H_0(\v{k}) + H_{ext}(t), 
\label{eqn:hamiltonian}
\\
H_0(\v{k}) &=& v (k_x \sigma_y - k_y \sigma_x ), \\
H_{ext}(t) &=& V\Theta(t-t_0) (\mathrm{a}_{x}(t) \sigma_y - \mathrm{a}_{y}(t) \sigma_x ).
\eeq
$H_{ext}(t)$ describes the external radiation source. The electrons are minimally coupled  
by the Peierls substitution $\v{k}\to \v{k} + e \v{A}(t)$,  where $\v{A}(t)=A_0 \v{a}(t)$ is the 
vector potential. We have set $\hbar=1$, $c=1$,  
the scalar potential to zero and ignored small magnetic effects.
Two polarizations are considered $\v{a}(t)= (\pm\cos\Omega t, \sin\Omega t)$ (circular) and 
$\v{a}(t)= (\cos\Omega t,0)$ (linear), where $T=2\pi/\Omega$ is the period of the external perturbation. 
The energy scale of the perturbation is given by $V= e v A_0 =e v E_0/\Omega$ where 
$E_0$ is the amplitude of the electric field and $v$ the speed of Dirac fermions. 
The dimensionless coupling constant $V/\Omega$ characterizes the strength of the perturbation. 
We assume the photon energies are small compared with the 
bulk gap of the TI ($<300$ meV).

The evolution operator, which is a $2\times 2$ matrix, 
obeys the time-dependent Schr\"odinger equation 
$i \partial_t U_{\v{k}}(t,t') = H(\v{k},t)U_{\v{k}}(t,t')$, with initial condition 
$U_{\v{k}}(t,t)=1$.  
Once the evolution operator is known, all other correlators can be calculated in terms 
of the initial state of the system. We first compute the retarded Green function in terms of the 
evolution operator using the equations of motion 
for the $c_{\v{k},\alpha}$ fields yielding the expression
$g^{r}_{\alpha\beta}(\v{k},t,t') \equiv -i \Theta(t-t')\langle \{c_{\v{k}\alpha}(t),c_{\v{k}\beta}^{\dagger}(t')\} \rangle
= -i \Theta(t-t') U_{\v{k}\alpha\beta}(t,t')$,
where $U_{\v{k}\alpha\beta}$ is the $(\alpha,\beta)$ matrix element of $U_{\v{k}}$. We can similarly compute 
the non-equilibrium electron distribution from the two-time lesser Green function 
which in terms of the evolution operator would read
as $g^{<}_{\alpha\beta}(\v{k},t,t')\equiv i\langle c_{\v{k}\beta}^{\dagger}(t') c_{\v{k}\alpha}(t)\rangle 
= i U_{\v{k}\gamma\beta}^{\dagger}(t',t_0)\langle c_{\v{k}\gamma}^{\dagger}(t_0)c_{\v{k}\delta}(t_0) \rangle U_{\v{k}\alpha \delta}(t,t_0)$
(summation over repeated indices is implied and greek indices take values $\{1,2\}$).  
We now make the simplifying assumption that a quasi-steady state has been reached, or equivalently that 
all correlations due to the initial state of the system had been washed away (mathematically we set 
$t_0 = -\infty$). Then the form of the evolution operator can be obtained analytically. 
Indeed, the evolution operator is $T$-periodic $U_{\v{k}}(t+T,t'+T)=U_{\v{k}}(t,t')$ and hence the 
retarded Green function also becomes $T$-periodic in the average time variable $\bar{t}=(t+t')/2$. 
This periodic structure allows for a simple analytical expression of the Wigner representation\cite{Kadanoff1989} 
of the retarded Green function. Here we focus on the pole structure of the non-equilibrium retarded Green function 
in the Wigner representation which in a sence is similar to the Lehmann representation of equilibrium correlators. 

\section{Spectral function}
\label{sec:spectral_function}
\begin{figure}[]
\subfigure{\includegraphics[width=0.23\textwidth]{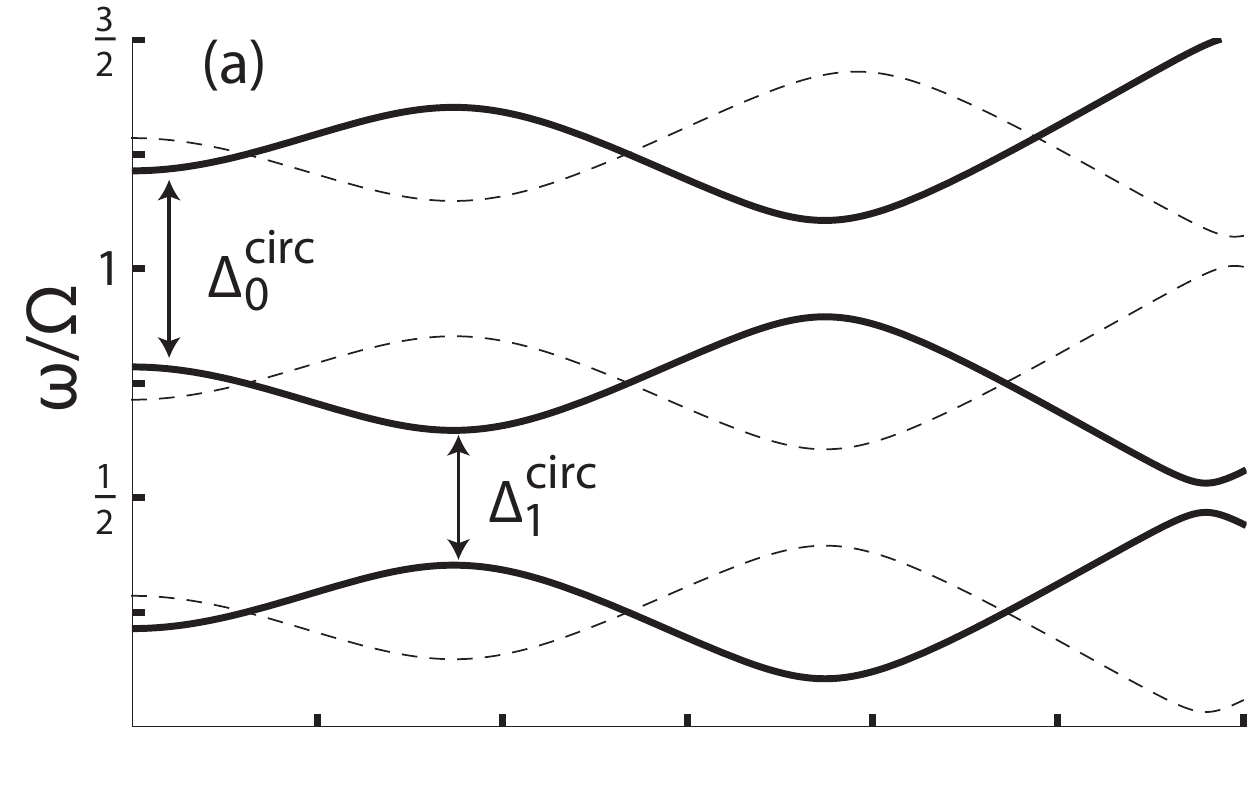}}
\subfigure{\includegraphics[width=0.21\textwidth]{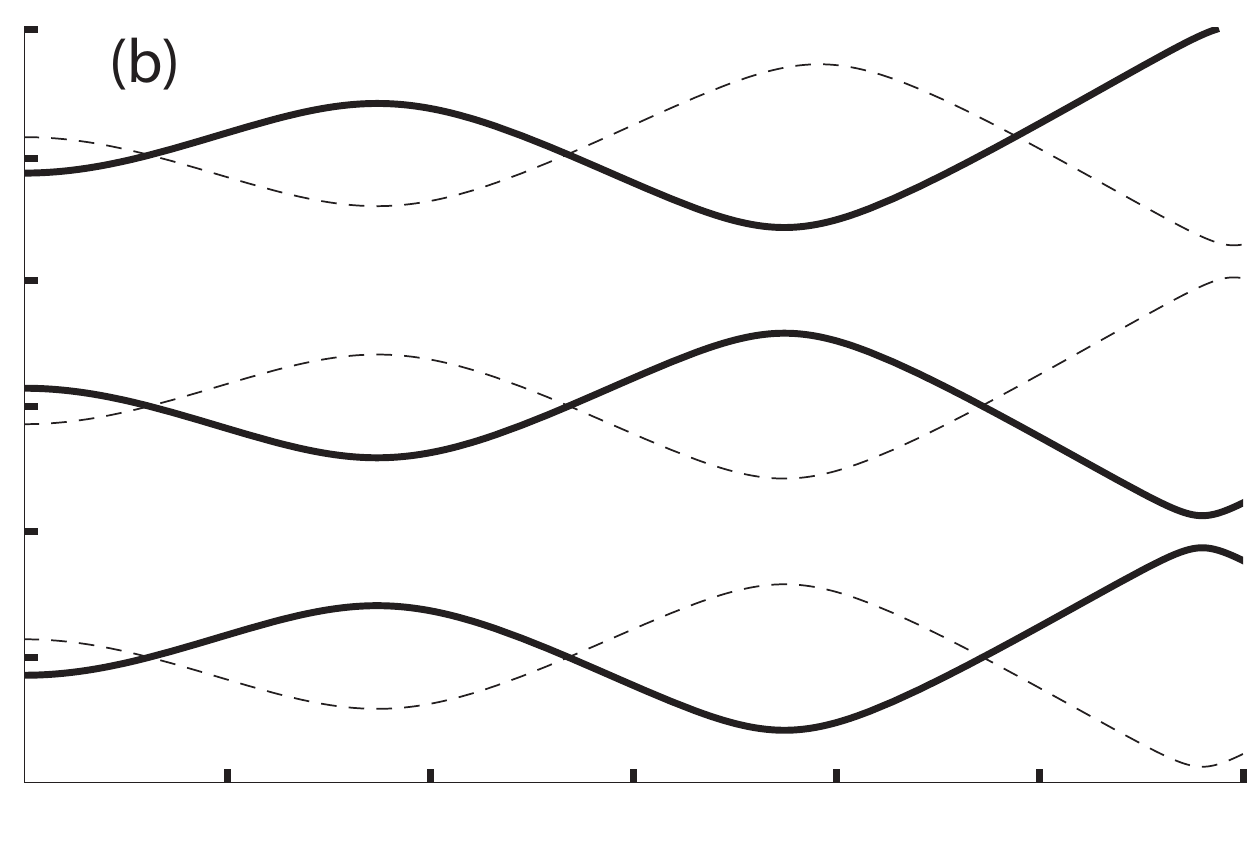}}
\subfigure{\includegraphics[width=0.23\textwidth]{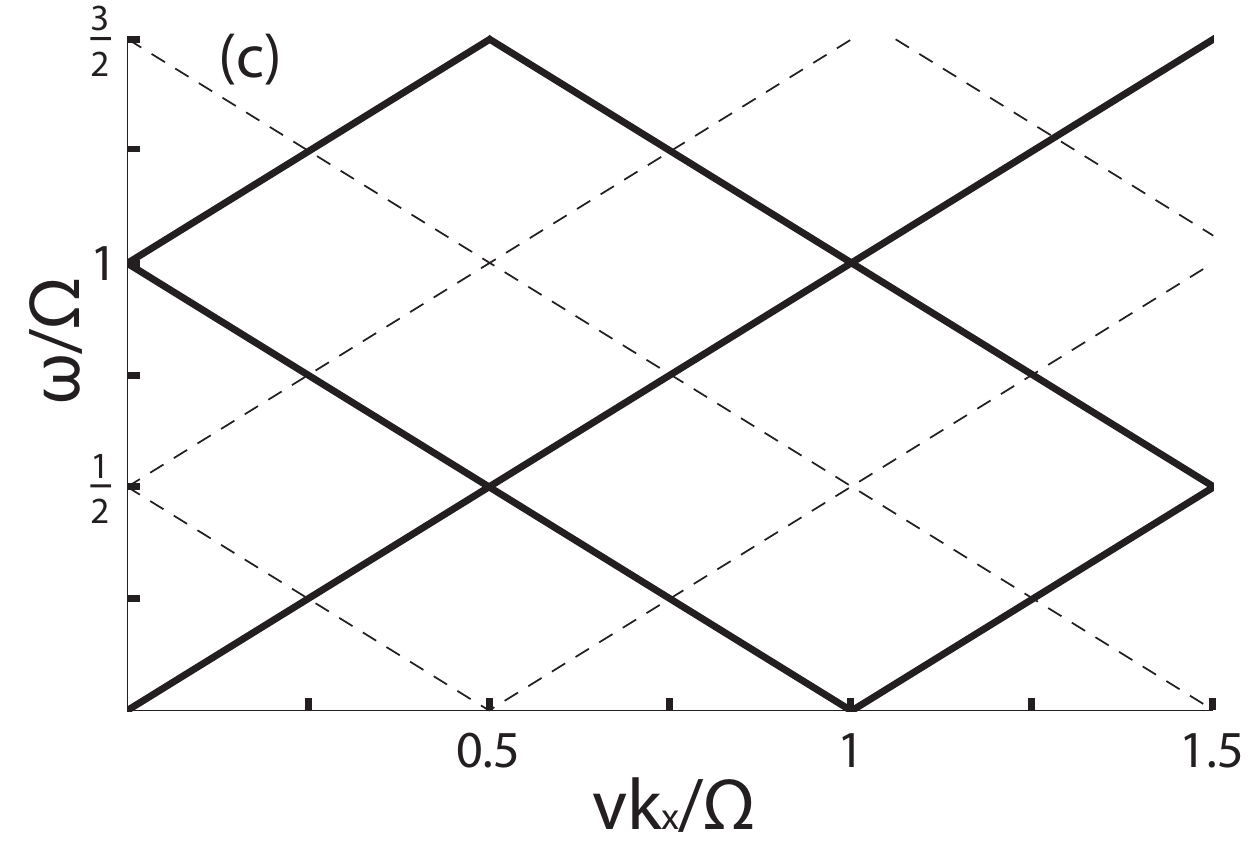}}
\subfigure{\includegraphics[width=0.21\textwidth]{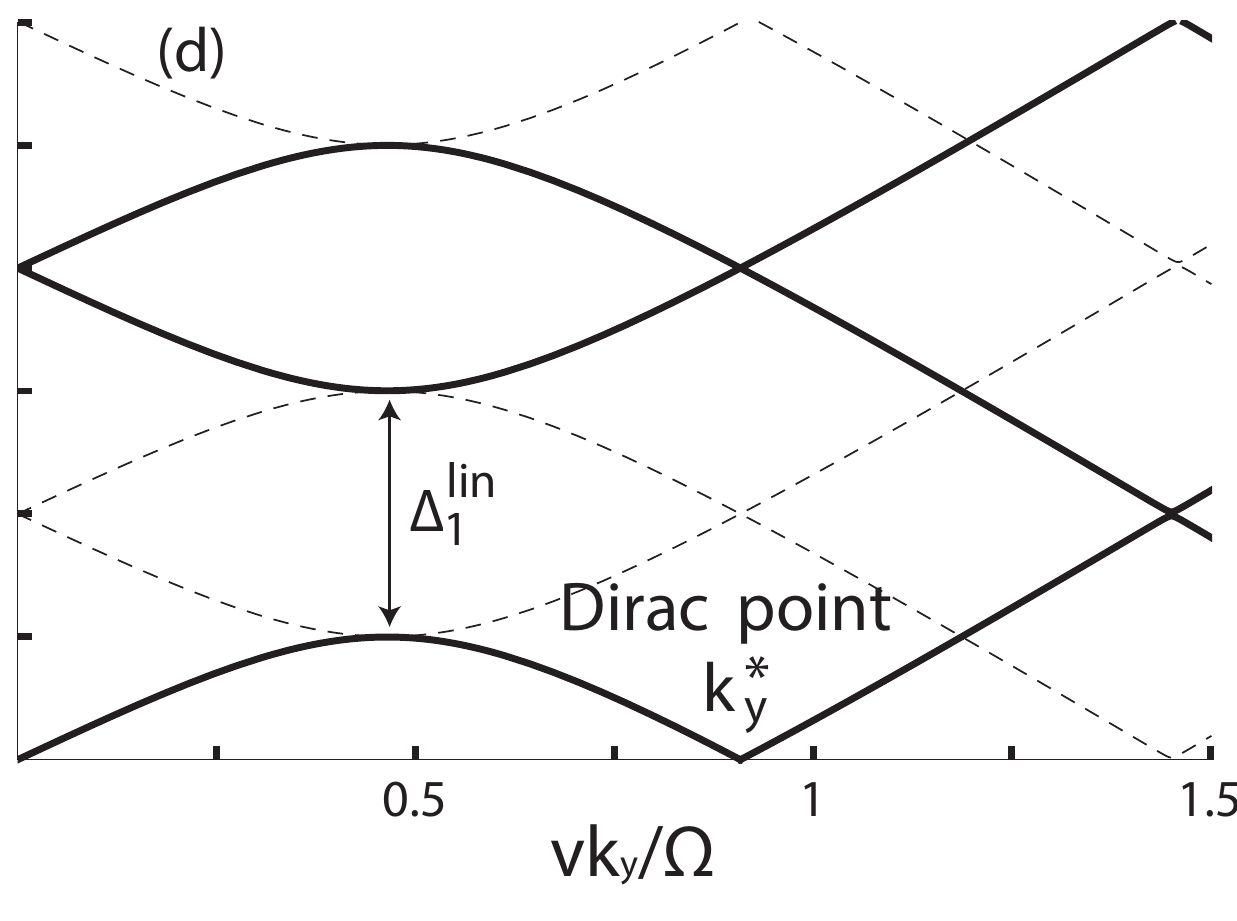}}
\caption{Spectrum calculated using an effective Floquet Hamiltonian truncated to six modes for 
circular [(a),(b)] and linear [(c),(d)] polarization, with momentum along
$k_x$ [(a),(c)] and along $k_y$ [(b),(d)]. The gap structure is clearly visible.  
Dashed (Solid) lines are periodic (static) bands with photon index $n$ 
odd (even). Here we take $V/\Omega=0.52$ which can be achieved experimentally. 
Note the Bloch-Siegert shifts in (a), (b), and (d).}
 \label{fig:analytic_spectra}
\end{figure}

It can be shown that 
the evolution operator\cite{Shirley1965} $U_{\alpha\beta}(t,t') = \sum_{n m\gamma} \left\langle \alpha n|\phi_{\gamma}^{m}\right\rangle \left\langle 
\phi_{\gamma}^{m}| \beta 0\right\rangle e^{-i \epsilon_{\gamma m}(t-t') + in\Omega t}$, satisfies the Schr\"odinger equation
where $\left\langle \alpha n|\phi_{\gamma}^{m}\right\rangle$ represents the ($\alpha, n$) component of
the eigenvector $|\phi_{\gamma}^{m} \rangle$ of the Floquet Hamiltonian (see Appendix \ref{app:floquet_theory}).
Hence Fourier transforming in $\bar{t}$ and the relative time $t-t'$ the
Wigner representation of the retarded Green function is,
\beq
g^r_{\alpha\beta}(\v{k},n,\omega) = \sum_{\gamma m} \frac{\langle \alpha n|\phi_{\v{k}\gamma}^{m}\rangle \langle 
\phi_{\v{k}\gamma}^{m}| \beta 0\rangle}{\omega - \epsilon_{\v{k}\gamma m} + n\Omega/2 + i 0^{+}}.
\label{eqn:wigner_rep_greenfunction}
\eeq
One can similarly obtain simple expressions for the inverse of the retarded Green function by 
noting that $U_{\v{k}}(t,t')^{-1}=U_{\v{k}}(t',t)$. Driven systems on a lattice with one-band 
have been studied before\cite{Turkowski2007,Freericks2008,Tsuji2008}.
The important point is that single-particle excitations, which occur at the poles of the imaginary part
of the retarded Green function are given by the quasienergies of the driven Dirac Hamiltonian, Eq.~\ref{eqn:hamiltonian}.
The index $n$ in Eq.~\ref{eqn:wigner_rep_greenfunction} 
represents the number of photons interacting with the Dirac electron.  
For example, a one-photon resonant transition creates exited states shifted by 
$\pm\Omega/2$ with respect to the original Dirac bands $\pm v k$ (see appendix \ref{app:green_func_PT}).
This dependence on $n$ was not considered in Ref.~\onlinecite{Kitagawa2011}. 

We are interested in the single-particle excited states which are given by the 
singularities of the non-equilibrium spectral function. In the Wigner representation
it is given by
\begin{align}
A(\v{k},n,\omega) &= -2 \textrm{Im} \textrm{Tr}[ g^{r}(\v{k},n,\omega)] \nn \\
&= -2 \textrm{Im} \bigg[\sum_{\gamma m} \frac{\langle \alpha n|\phi_{\v{k}\gamma}^{m}\rangle \langle 
\phi_{\v{k}\gamma}^{m}| \alpha 0\rangle}{\omega - \epsilon_{\v{k}\gamma m} + n\Omega/2 + i 0^{+}} \bigg].
\label{eqn:spectraldef}
\end{align}
Of particular interest is the average over a period $T$ of the driving force
which is just the $n=0$ term,
\beq
A(\v{k},0,\omega) = 2\pi \sum_{\alpha \gamma m} |\langle \alpha 0|\phi_{\v{k}\gamma}^{m} \rangle|^2 \delta(\omega - \epsilon_{\v{k}\gamma m}).
\label{eq:avg_spectral}
\eeq
Using  the completeness of the 
$|\phi_{\v{k}\gamma}^{m} \rangle$ states we can verify 
that it satisfies the sum rule $\int (d\omega/2\pi) A(\v{k},0,\omega) = 2$.
Such property is not shared by any other moment of the non-equilibrium spectral function.
In equilibrium, the spectral function
does satisfy this sum rule which, in that case, derives from fermion conservation and the 
factor of two comes from the spin. In this sense, the average calculated above 
is more physical than the non-equilibrium spectral function, by itself.

We now consider the general structure of $\epsilon_{\v{k}\alpha n}$.
See appendix \ref{app:floquet_theory} and similar spectra obtained 
for irradiated graphene\cite{Syzranov2008,Oka2009,Zhou2011}.
One can usually find an approximate form of the quasienergies from  
a truncated Floquet Hamiltonian,
$\langle \alpha n|\mathcal{H}_F|\beta m \rangle = H_{\alpha\beta}^{n-m} + n\Omega \delta_{\alpha\beta}\delta_{n m}$. 
For example, in Fig.~\ref{fig:analytic_spectra} we show the quasienergies 
for circular and linear polarizations with six modes as
a function of momentum along $k_x$ and $k_y$ (see also Fig.~\ref{fig:quasienergy_kx_ky}). 
We verified that higher modes do not change the 
spectrum significantly in the range of energies we consider.
One can understand the structure of the spectrum as composed of copies of 
the original Dirac bands shifted by multiples of $\Omega$, i.e., $\epsilon^{0}_{\v{k} 1 n} = v k + n\Omega$,
$\epsilon^{0}_{\v{k} 2 m} = -v k + m\Omega$, and treating the effects
of nonzero $V$ perturbatively at the crossings~\cite{Shirley1965}.
If there is a non-zero coupling, the bands exhibit an anti-crossing (avoided crossing).
For $V=0$ note that $\mathcal{H}_{F}$ has time-reversal invariance and obviously  
time-translation invariance. We will see how these symmetries will be explicitly broken by the 
perturbation. 
In general, band crossings are associated with symmetries 
of the system. If there are no symmetries, any crossings/degeneracies
are accidental. An early result of Von-Neumann and Wigner for time-independent Hamiltonians  
established that two (three) parameters are necessary to produce an accidental degeneracy 
for real (complex) Hamiltonians. Hence by varying only one parameter, such as $k_x$ or $k_y$,
we expect to produce only avoided crossings. 

\begin{figure}[]
\subfigure{\includegraphics[width=0.23\textwidth]{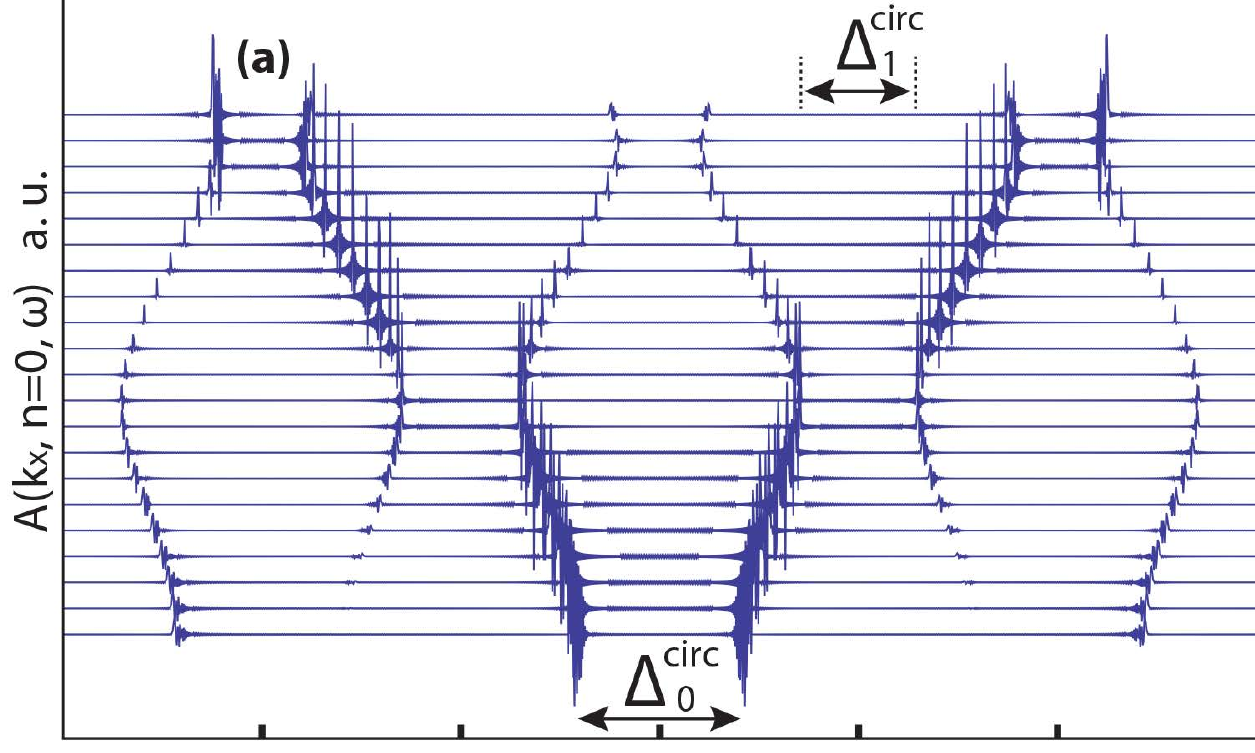}}
\subfigure{\includegraphics[width=0.23\textwidth]{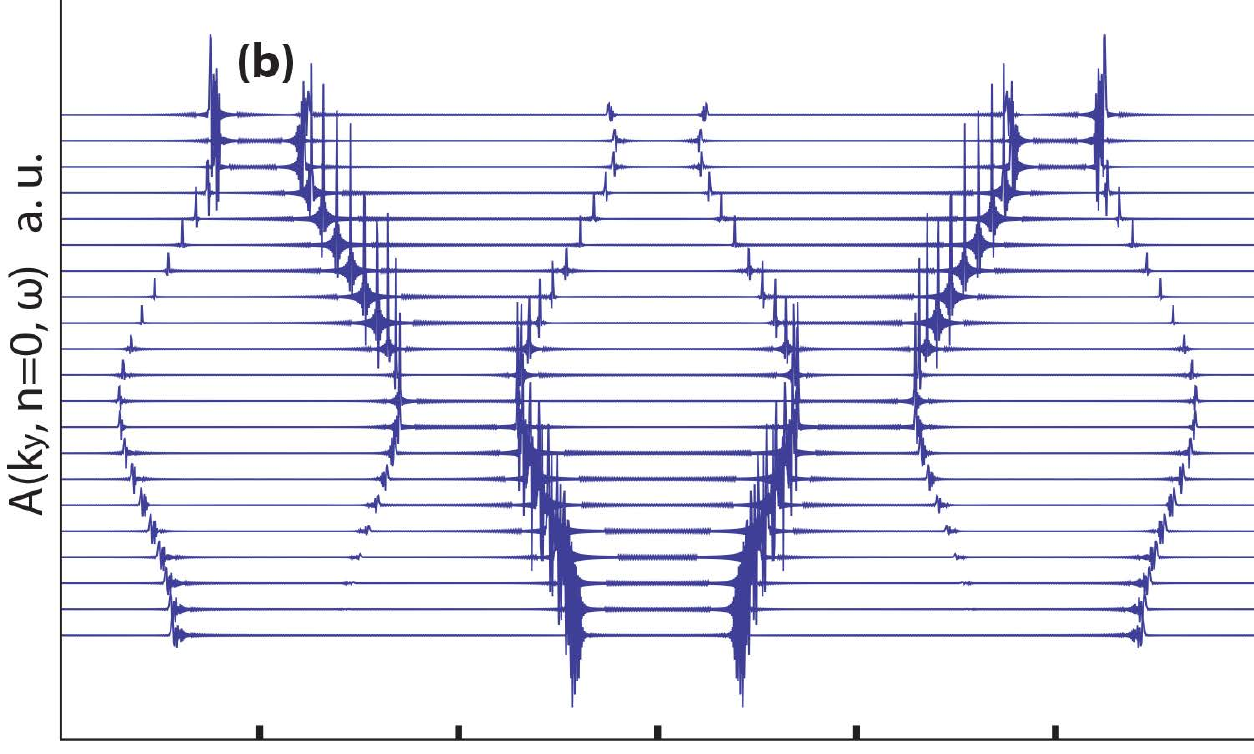}}
\subfigure{\includegraphics[width=0.23\textwidth]{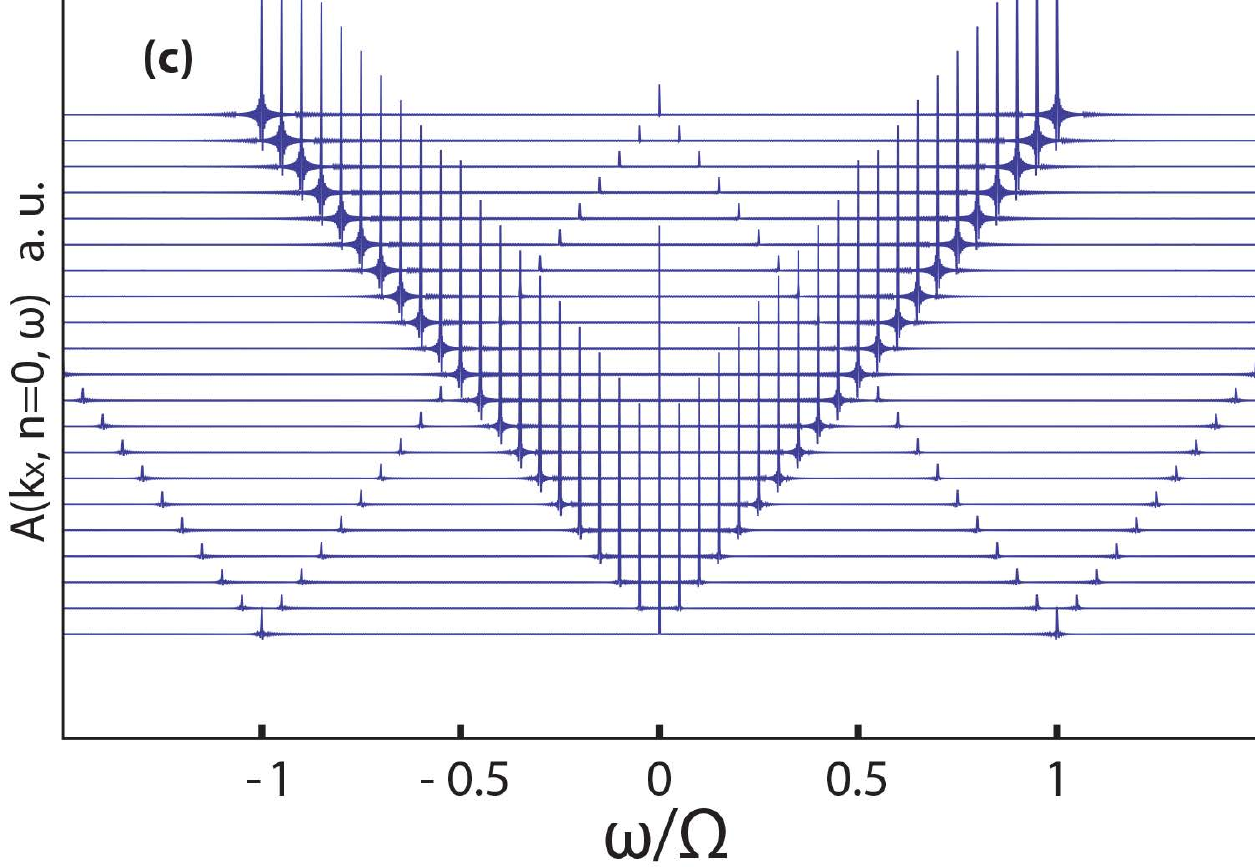}}
\subfigure{\includegraphics[width=0.23\textwidth]{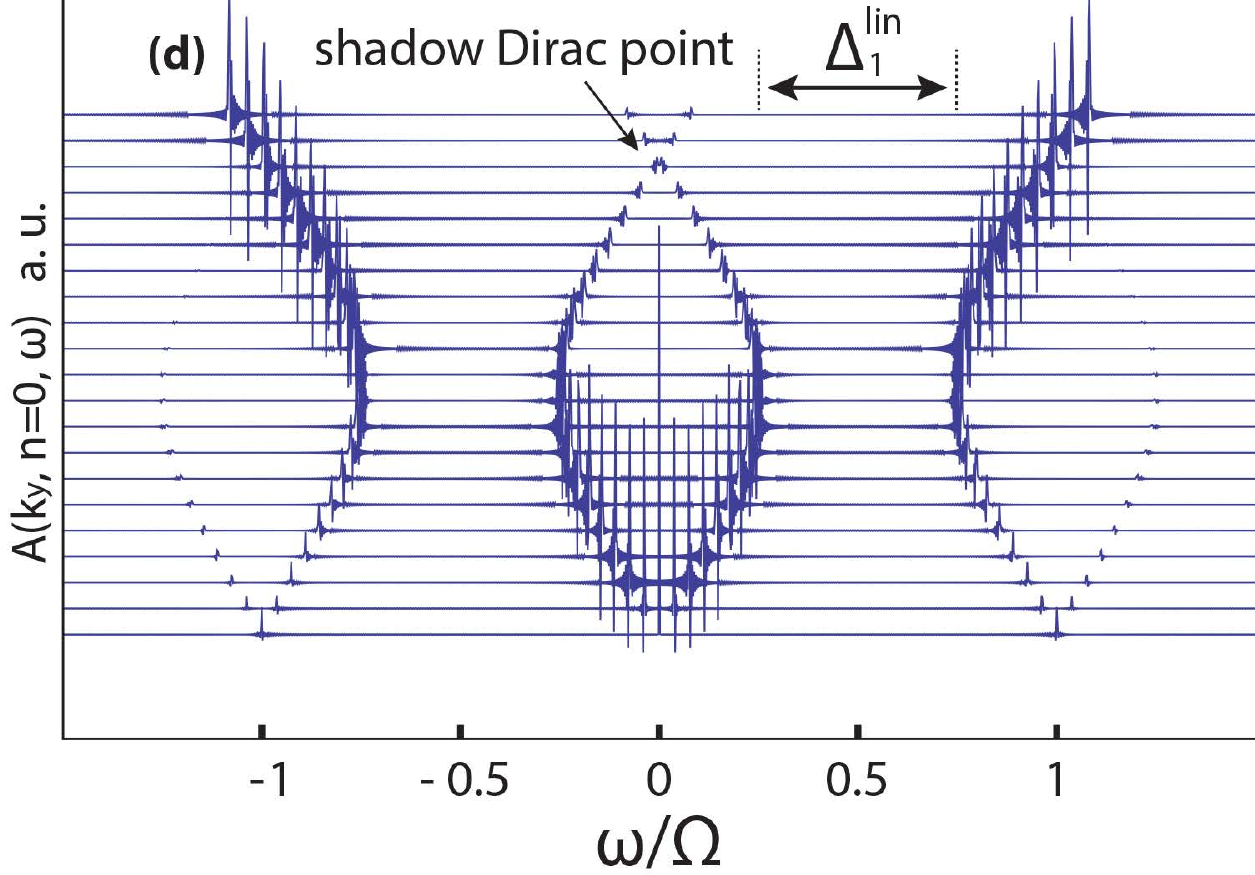}}
\caption{(Color online) Average of the spectral function, $A(\v{k},n=0,\omega)$, from numerically evolving the evolution 
operator in time, Fourier transforming and taking the imaginary part of the retarded Green function.
We consider circular [(a), (b)] and linear [(c), (d)] polarization, with momentum along the $k_x$-axis [(a), (c)] and $k_y$-axis [(b), (d)].
In all panels, the horizontal lines correspond to different values of 
$k_x$ ($k_y$) from $k_x=0$ ($k_y=0$) to  $v k_x = \Omega$ ($v k_y = \Omega$) 
in steps of $0.05$.
We see the photoinduced gaps at the Dirac point and finite momentum. An exact
crossing occurs at $k_y^{*}$ with linear polarization, panel (d), 
giving rise to a shadow Dirac point. For $V/\Omega =0.52$, the spectral 
weight is concentrated near the original Dirac bands but significant weight 
is still observed at shadow bands which are displaced by multiples of $\Omega$.
From the graphs we can read $\Delta_0^{circ}= 0.43\Omega$, $\Delta_{1}^{circ}= 0.31\Omega$
and $\Delta_{0}^{lin}=0$, $\Delta_{1}^{lin}=0.52\Omega$.}
\label{fig:num_spectra}
\end{figure}

Consider the case of circularly polarized photons. In this case 
time reversal symmetry is broken, and since no other symmetries remain 
we expect only avoided crossings in the spectrum [Figs.~\ref{fig:analytic_spectra}(a) and \ref{fig:analytic_spectra}(b)].
If the perturbation is small, $V/\Omega\ll 1$,
we can restrict the analysis to the two crossing bands in question. For concreteness, 
consider the crossing at $v k_x  \approx \Omega/2$ and $k_y=0$.
The effective Hamiltonian is,
\beq
H_{2v k_x=\Omega}=
\left( 
\begin{array}{cc} 
H_0 +\Omega & iV \sigma^{-} \\ 
- iV \sigma^{+} & H_0 
\end{array} 
\right),
\label{heff4}
\eeq
where we used   
$H_{ext}(t) = i V\sigma^{-} e^{i \Omega t} + h.c.$ and $\sigma^{\pm } =(\sigma_x \pm i\sigma_y)/2$. 
In the absence of the external perturbation, the four eigenvalues of the above matrix 
are $ vk_x, -v k_x, vk_x +\Omega$, and $-v k_x +\Omega$ corresponding to the 
eigenvalues of the diagonal terms. Note that for $V=0$  
two of these bands cross at $v k_x =\Omega/2$. The effect of a small non-zero $V/\Omega$ is
to open a gap in the spectrum at $v k_x= (\Omega/2)\sqrt{(2V^2 + \Omega^2)/(V^2+\Omega^2)}$ of 
magnitude
$\Delta_{1}^{circ}/\Omega = V/\sqrt{V^2+ \Omega^2}= V/\Omega + O(V^3)$. 
The magnitude of this gap agrees with previous perturbative calculations using the 
rotating wave approximation\cite{Syzranov2008}. For 
larger $V/\Omega$, where the above approximation of retaining just two
modes is not valid, the resonance occurs at momenta 
$v k_x < \Omega/2$ [see Figs.~\ref{fig:analytic_spectra}(a) and \ref{fig:analytic_spectra}(b)].
An exact numerical solution of the time-dependent Schr\"odinger equation confirms 
this result as shown in Figs.~\ref{fig:num_spectra}(a) and \ref{fig:num_spectra}(b). 
The Bloch-Siegert shift\cite{Dittrich1998} observed is an effect beyond the scope of
the rotating wave approximation.

Next we consider the Dirac point where a gap is photoinduced\cite{Oka2009}
due to the absorption and subsequent emission of a photon by an electron 
near the Dirac point. The effective Hamiltonian is   
\beq
H_{\v{k}= 0}=
\left( 
\begin{array}{ccc}
H_0+ \Omega & iV \sigma^{-} & 0 \\ 
-i V \sigma^{+} & H_0 & iV \sigma^{-} \\ 
0 & -i V \sigma^{+} & H_0 -\Omega 
\end{array} 
\right).
\label{heff6}
\eeq
In the limit of $V/\Omega \ll 1$ the gap is $\Delta_{0}^{circ}\approx 2(V^2/\Omega)$ (see appendix \ref{sec:gap_dirac_pertubation_theory}). 
Here we provide an alternative derivation, 
valid to all orders in perturbation theory, by noting that
at the Dirac point, the Hamiltonian in Eq.~(\ref{eqn:hamiltonian}) 
is one of the few analytically solvable driven two-level models\cite{Galitski2011,*Wilson2012}; formally equivalent to a 
spin-$1/2$ in a circularly polarized magnetic field. Explicitly, the evolution operator at $\v{k}=0$ is given 
by $U_{\v{k}=0}(t,t')= e^{-i \sigma_z\Omega t/2}e^{- i H (t-t')} e^{ i \sigma_z\Omega t'/2}$,
where $H = V\sigma_y - \Omega \sigma_z/2$.  
Hence the gap at the Dirac point is\cite{Oka2009} 

\beq
\Delta_{0}^{circ} = \sqrt{\Omega^2 + 4 V^2} ~~\textrm{mod}~\Omega.
\label{eq:gap_dirac}
\eeq

\section{Shadow Dirac points}
\label{sec:shadow_points}
In Figs.~\ref{fig:analytic_spectra}(c) and \ref{fig:analytic_spectra}(d), we see band crossings at $\v{k}=0$ 
and at finite momentum. In this section we show that these crossings are protected 
by a \textit{dynamical} symmetry of the system in the presence of light with linear polarization. 
The Floquet operator for polarization along $x$,  in the basis of eigenvectors of $H_0$ is
\begin{align}
\tilde{\mathcal{H}}_F^{lin} = v |\v{k}| \sigma_z +  V&\cos\varphi_{\v{k}} \cos(\Omega t) \sigma_z \nn  \\ & - V\sin\varphi_{\v{k}} \cos(\Omega t) \sigma_y -i \partial_t,
\end{align} 
where $\varphi_{\v{k}}$ denotes the angle between the 
momenta and and the $x$-axis.
If the momentum is along $k_x$, the external perturbation commutes with $H_0$ and produces the 
trivial spectrum shown in Fig.~\ref{fig:analytic_spectra}(c) with Dirac points at $v k_x = n\Omega$. 
If the momentum is perpendicular to the polarization the Hamiltonian is 
$\tilde{\mathcal{H}}^{lin}_F = v |k_y|\sigma_z - V \cos\Omega t~ \sigma_y - i \partial_t$ and
satisfies

\beq
\hat{P} \tilde{\mathcal{H}}_F^{lin} \hat{P}^{-1} = \tilde{\mathcal{H}}_F^{lin},
\eeq
where $\hat{P} = \sigma_z e^{(T/2)\partial_t}$ is the `parity' operator\cite{Grossmann1992}.
This operator shifts time by $t \to t+ T/2$ and flips the spin operator $\sigma_y \to -\sigma_y$. 
This means that eigenstates $\phi_{\alpha n}$ can be defined with good parity quantum number according to whether 
$\alpha+n$ is even or odd.
In other words, the external perturbation does not have matrix 
elements between states of different symmetries and hence crossings of these 
bands cannot be gapped; they are symmetry-protected.
This can be explicitly shown by writing $\tilde{\mathcal{H}}_F^{lin}$ in 
frequency space, in the basis of eigenvectors of $H_0$, and noting that it
splits into disjoint blocks
$\tilde{\mathcal{H}}_F^{lin} = \tilde{\mathcal{H}}_{even}\oplus \tilde{\mathcal{H}}_{odd}$ (see appendix \ref{app:symetry_protected_cross}).  
One implication is that the $\v{k}=0$ point remains gapless
\beq
\Delta_{0}^{lin}= 0,
\eeq
to all orders in perturbation theory, as the bands $\epsilon_{\v{k}1,0},\epsilon_{\v{k}2,0}$  
have different parities. Indeed, at this point the Floquet operator vanishes for linear plarization. 

Similarly, there are symmetry-protected band crossings at finite momentum between the bands  
$\epsilon_{\v{k}1,odd}$ and $\epsilon_{\v{k}2,odd}$. This is made explicit in 
Fig.~\ref{fig:analytic_spectra}(d) where the bands
$\epsilon_{\v{k}1,-1}$ and $\epsilon_{\v{k}2,1}$ cross at  $\pm k_y^{*}$.
Including $n=0,\pm 1,\pm 2$ Fourier modes we obtain zero-energy eigenvalues of 
 $\mathcal{H}_F^{lin}$ at momenta
$v k_y^{*}/\Omega = \big[10- 2 (V/\Omega)^2 - \sqrt{(V/\Omega)^4+ 8 (V/\Omega)^2 +36}\big]^{1/2}/2$.
This expression is accurate to $O(V^2)$,
\beq
\frac{v k_y^{*}}{\Omega} = 1- \frac{V^2}{3\Omega^2} + O(V^4).
\eeq
For $V/\Omega=0.52$ we have $v k_y^{*}/\Omega \approx 0.91$.  
As momentum increases, with fixed $V/\Omega<1$, the crossings asymptotically 
move\cite{Grossmann1992} towards $v k_y^{*} =n \Omega$.
On the other hand, the crossing of the $\epsilon_{\v{k}1,0}$ and $\epsilon_{\v{k}2,1}$ bands can be 
gapped as these bands belong to the same symmetry class. The magnitude of this
gap is $\Delta_1^{lin}= V$ for $V/\Omega \ll 1$.
For arbitrary direction in 
momentum space, other than $k_x$ and $k_y$, the states have no well defined parity,   
degeneracies are not symmetry-protected, and gaps develop in the spectrum (see Fig.~\ref{fig:quasienergy_kx_ky}).
The above considerations show that band touchings occur only at these 
special points in momentum space and that the
existence and position of these points can be  
\textit{engineered} with a properly chosen frequency.

\section{Discussion and conclusion}
\label{sec:dis_conclu}
In a TrARPES experiment, the measured photo-current is proportional to 
the two-time (non-equilibrium) lesser Green function\cite{Freericks2009} which in turn 
is proportional to the distribution function (generally unknown) and the 
spectral function. 
Furthermore, for a system were there is no phase coherence between the pump and probe pulses,
we expect the photocurrent to time-average over the period of the driving force.
These arguments suggest that 
the measured spectrum would be characterized qualitatively by the average of 
the spectral function, Eq.~\ref{eq:avg_spectral} (calculated numerically in 
Fig.~\ref{fig:num_spectra}).
For concreteness, if the electric field is $E_0 \approx 2.2\times 10^7$ 
V/m and the photon energies are $120$ meV then using $v=5\times 10^5$ m/s as the speed of Dirac electrons on the surface of 
Bi$_2$Se$_3$, we obtain a coupling $V/\Omega=0.52$. Using  Eq.~\ref{eq:gap_dirac}, we obtain a gap 
$\Delta_0^{circ} = 51$ meV, in agreement with 
our simulation in Fig.~\ref{fig:num_spectra}a and the experiment\cite{Wang}, $\Delta_0^{circ} = 53 \pm 4$ meV. At 
finite momentum and circular polarization, we obtain $\Delta_{1}^{circ}=37$ meV
along $k_x$ and $k_y$, and so the spectrum is isotropic (see Fig.~\ref{fig:num_spectra}a,b). 
For linear light, we obtain $\Delta_{0}^{lin} =0$ along $k_x$ and $k_y$ (Fig.~\ref{fig:num_spectra}c,d).
At finite momenta, $\Delta_{1}^{lin} =0$ along $k_x$ but $\Delta_{1}^{lin} =62$ meV
at $vk_y\approx \Omega/2$, in agreement with the experiment\cite{Wang}  
 $\Delta_{1}^{lin} =62 \pm 5$ meV. The position of the first shadow Dirac point is 
$v k_y^{*}= 109 $ meV. 

In conclusion, we have calculated the non-equilibrium spectral function
of electrons at the surface of TIs in the presence of an incident light with circular and 
linear polarization. Depending on polarization the system is an anisotropic metal with
multiple Dirac cones or an insulator. This theory along with the experimental 
technique would allows for optical engineering of non-equilibrium spectra in topological materials.

\textit{Acknowledgments}.
We thank Jim Freericks, Kostya Kechedzhi, Stefan Natu and Lev Bishop for discussions.
YHW and NG would like to acknowledge support from Department of Energy
Office of Basic Energy Sciences Grant No. DE-FG02-08ER46521, BMF the 
NSF through the PFC@JQI and Conacyt and VG DOE-BES (DESC$0001911$).  

\textit{Note added}. After completion of this work we learned
about Ref.~\onlinecite{Delplace} which contains partial overlap with our work.

\appendix
\section{Periodically driven two-level Hamiltonians}
\label{app:floquet_theory}
Here we provide a brief review of periodically driven two-level systems\cite{Shirley1965}.
If the Hamiltonian is periodic in time $H(\v{k},t+T)  = H(\v{k},t)$, with period $T$,
the solution of the Schr\"odinger equation   
\beq
i \partial_t \psi_{\v{k}}(t) &=& H(\v{k},t) \psi_{\v{k}}(t),
\eeq
can always be written as 
\beq
\psi_{\v{k}}(t) = \phi_{\v{k}}(t) e^{-i \epsilon_{\v{k}} t},
\eeq
where $\phi_{\v{k}}(t)=\phi_{\v{k}}(t+T)$ is periodic and the phase $\epsilon_{\v{k}}$ is 
the \textit{quasienergy}, which is defined modulo $\Omega=2\pi/T$. Substituting into
the Schr\"odinger equation gives the eigenvalue problem 
\begin{align}
\mathcal{H}_F\phi_{\v{k}\gamma}(t) \equiv \left(H(\v{k},t) - i\partial_t\right)\phi_{\v{k}\gamma}(t) =\epsilon_{\v{k}\gamma} \phi_{\v{k}\gamma}(t), 
\label{eq:eigenvalue_floquet}
\end{align}
where the $\gamma=\{1,2\}$  distinguishes distinct eigenstates of the \textit{Floquet} Hamiltonian $\mathcal{H}_F$
Defining $\phi_{\v{k}\gamma}(t) = \sum_{n} \phi_{\v{k}\gamma}^{n} e^{in\Omega t}$
and $H(\v{k},t) = \sum_{n} H^{n}(\v{k}) e^{in\Omega t}$ we obtain the frequency representation of 
Eq.~\ref{eq:eigenvalue_floquet},
\begin{align}
\sum_{m\beta} \langle \alpha n|\mathcal{H}_F|\beta m \rangle 
\left\langle \beta m|\phi_{\gamma}^{l}\right\rangle = \epsilon_{\gamma l}\left\langle \alpha n|\phi_{\gamma}^{l}\right\rangle,
\end{align}
where  
$\langle \alpha n|\mathcal{H}_F|\beta m \rangle = H_{\alpha\beta}^{n-m} + n\Omega \delta_{\alpha\beta}\delta_{n m}$.
We omit the momentum label to simplify expressions when there is no danger of confusion.
The quasienergies are $\epsilon_{\alpha n} = \epsilon_{\alpha} + n \Omega$, where $n$ is an integer. 
For circular polarization, the external driving can be written as 
$H_{ext}(t)=i V \sigma^{-} e^{i \Omega t}- i V \sigma^{+} e^{-i \Omega t}$,
where $\sigma^{\pm} =(\sigma_x  \pm i \sigma_y)/2$. Hence the
Floquet matrix corresponding to circular polarization is
$[\mathcal{H}_F]_{nm} = \delta_{n,m} H_0 + i V \sigma^{-}\delta_{n,m-1} - i V \sigma^{+}\delta_{n,m+1} - \delta_{nm} n\Omega$
or explicitely,
\begin{align}
\mathcal{H}_{F}=
\left( 
\begin{array}{ccccccc}
\ddots & & & & \\ 
 & & & & \\
& & H_0+\Omega & i V \sigma^{-} & 0 & &\\ 
& & - i V \sigma^{+} & H_0 & i V \sigma^{-} & &\\
& & 0 & -i V \sigma^{+} & H_0 -\Omega & &\\
& & & & &  &  \\
& & & & &  &\ddots
\end{array} 
\right),
\label{HF}
\end{align}
Were $H_0$ is the unperturbed Dirac Hamiltonian. 
For linear drive $\v{A}(t)=A_0(\cos\Omega t,0)$ and the external drive is 
$H_{ext}(t) = V \sigma_y \cos\Omega t$ which leads to 
\begin{align}
\mathcal{H}_{F}^{lin}=
\left( 
\begin{array}{ccccccc}
\ddots & & & & \\ 
 & & & & \\
& & H_0+\Omega &  V \sigma_y/2 & 0 & &\\ 
& &  V \sigma_y/2 & H_0 &  V \sigma_y/2 & &\\
& & 0 &  V \sigma_y/2 & H_0 -\Omega & &\\
& & & & &  &  \\
& & & & &  &\ddots
\end{array} 
\right),
\label{HFlinear}
\end{align}
In Fig.~\ref{fig:analytic_spectra} we have truncated the Floquet matrix 
to six Fourier modes and obtained the spectrum of the driven system for 
circular and linear polarization.

\section{Dynamical symmetry of the system}
\label{app:symetry_protected_cross}
\begin{figure}[]
\subfigure{\includegraphics[width=0.30\textwidth]{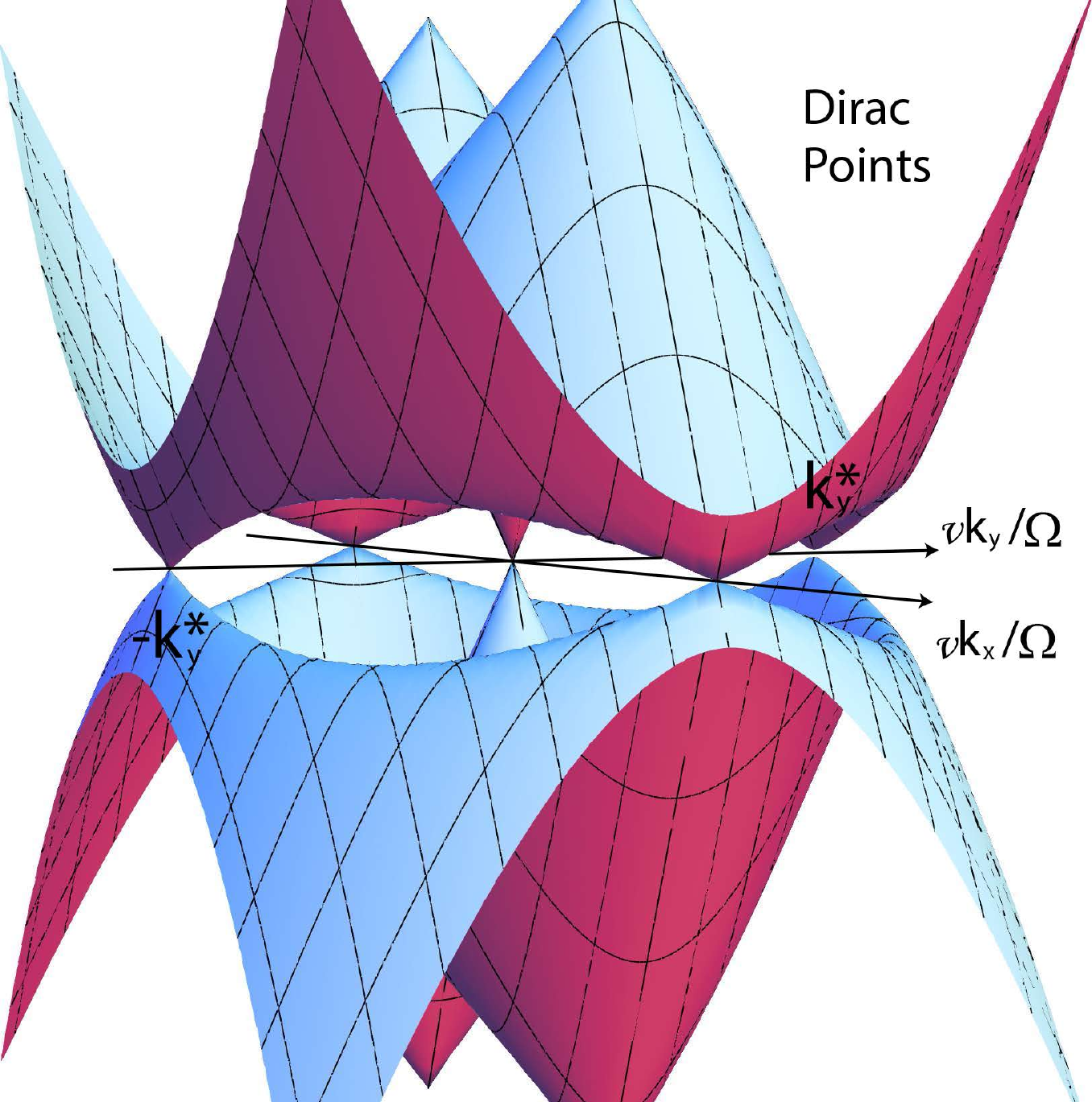}}
\caption{(Color online) The lowest branch of the quasienergies, $\pm \epsilon_{\v{k}1,0}$ for linear 
polarization. 
The original Dirac cone and the anisotropic shadow Dirac cones are clearly visible.
Other Dirac points can be seen in higher branches. The parameters are the same as those used in Figs.~\ref{fig:analytic_spectra}(c) and ~\ref{fig:analytic_spectra}(d) }
\label{fig:quasienergy_kx_ky}.
\end{figure}

In the basis of the vectors
$\{ (1,~ i e^{i\varphi_k})^{T}/\sqrt{2}, (-1,~ i e^{i\varphi_k})^{T}/\sqrt{2} \}$  
the unperturbed Hamiltonian is diagonal $\tilde{H}_0= v |\v{k}|\sigma_z$ and for linear polarization the 
perturbation takes the form  
$\tilde{H}_{ext}(t) = V\cos\varphi_\v{k} \cos\Omega t\sigma_z - V\sin\varphi_\v{k} \cos\Omega t \sigma_y$, 
where $\varphi_{\v{k}}$ is the angle of the electron momentum with respect to the polarization which is taken to define the $x$-axis. 
For the case of momenta perpendicular to the polarization the Floquet operator is  
$\tilde{\mathcal{H}}^{lin}_F = v |k_y|\sigma_z - V \cos\Omega t~ \sigma_y - i \partial_t$ and its
frequency representation is 
$[\tilde{\mathcal{H}}^{lin}_F]_{nm} = v |k_y| \delta_{nm}\sigma_z - V(\delta_{n,m-1} +\delta_{n,m+1})\sigma_y/2 + n\Omega\delta_{nm}$
or explicitly shown in Eq.~(\ref{HFlinear}).
Note that $\tilde{\mathcal{H}}^{lin}_F$ can be divided into two disconnected blocks (symmetry classes). 
For example, the states $v |k_y|-\Omega$ and $-v |k_y|+\Omega$ belong to
distinct blocks with no matrix element connecting them to any order in perturbation theory 
and hence they cross. Similarly, the branches $v |k_y|$ and $-v |k_y|$ do not 
mix to any order in perturbation theory and hence the crossing at $\v{k}=0$ is also symmetry-protected as 
concluded in the main text. Finally we note that an electron in the state $\pm v |k_y|$ always changes 
chirality upon interacting with a linearly polarized photon leaving it in the state $\mp v |k_y|$. This is in contrast
to circular polarized photons where there is a finite probability of leaving the electron 
with the same chirality.
\begin{widetext}
\beq
\tilde{\mathcal{H}}_{F}^{lin}=
\left( 
\begin{array}{cccccccc}
\ddots&&&&&&&\\
&v |k_y| + \Omega     &      0           &      0 &    i V/2 & 0             & 0               & \\ 
&0                    & - v |k_y|+\Omega & - iV/2 &        0 & 0             & 0               &\\ 
&0                    &           - iV/2 & v|k_y| &        0 & 0             & iV/2            &\\ 
&        -iV/2        &               0  &      0 & -v |k_y| & - iV/2        & 0               &\\
&        0            &               0  &      0 &    i V/2 & v|k_y|-\Omega & 0               &\\
&        0            &               0  &  -iV/2 &        0 &             0 & -v |k_y|-\Omega & \\
&&&&&&&\ddots
\end{array} 
\right).
\label{HFlinear}
\eeq
\end{widetext}
In Fig.~\ref{fig:quasienergy_kx_ky} we have calculated numerically the lowest branch of quasienergies
with a truncated Floquet Hamiltonian to six modes as a function 
of $\v{k}$ for linear polarization. Note that only one
branch is independent due to the constrain $\epsilon_{\v{k}1,0} +\epsilon_{\v{k}2,0}=0$. We have 
verified numerically the linearity of the dispersion near the band touchings.  

\section{Gap at the Dirac point from perturbation theory}
\label{sec:gap_dirac_pertubation_theory}
In the presence of circularly polarized light a gap at $\v{k}=0$ will develop. It can be understood 
intuitively as arising 
from renormalization effects due to virtual interactions between the branches $vk$ and $vk \pm \Omega$.
To see this let us consider the truncated Hamiltonian shown in Eq.~(\ref{heff6}) 
which contains three branches with Fourier modes  $n=0,\pm 1$.
Direct diagonalization involves solving an equation of sixth degree. 
To reveal the nature of the gap we proceed in a different way. We are only interested in 
the renormalization of the $n=0$ mode corresponding to the bands $\pm v|\v{k}|$ 
 near $\v{k}=0$. The set of equations to solve is (omitting momentum label)

\beq
(H_0 + \Omega) \phi^{1}_{\alpha} + i V \sigma^{-}\phi^{0}_{\alpha} &=& \epsilon_\alpha \phi^{1}_{\alpha}, \\
- i V \sigma^{+} \phi_{\alpha}^{1} + H_0 \phi_{\alpha}^{0} + i V \sigma^{-}\phi^{-1}_{\alpha} &=& \epsilon_\alpha \phi_{\alpha}^{0},
\label{eqn:0mode}
\\
-i V \sigma^{+} \phi_{\alpha}^{0} + (H_0- \Omega)\phi_{\alpha}^{-1} &=& \epsilon_\alpha \phi_{\alpha}^{-1}.
\eeq
If we assume that $\Omega >> \epsilon_\alpha$ then from the first and third equations 
we solve for $\phi_{\alpha}^{\pm 1}$ as
$\phi^{\pm 1}_{\alpha} = -i(V/\Omega) \sigma^{\mp } \phi^{0}_{\alpha}$.
Substituting back into Eq.~\ref{eqn:0mode} we obtain an effective equation for the   
$n=0$ state,
$(H_0 - V^2 \sigma_z/\Omega) \phi^{0}_{\alpha} = \epsilon_{\alpha} \phi_{\alpha}^{0}$.
The eigenvalues of the renormalized Hamiltonian are $\pm \sqrt{v^2 k^2 + (V^2/\Omega)^2}$ 
with a gap at $\v{k}=0$ of magnitude
$\Delta_{k=0}^{circ}/\Omega \approx 2 (V/\Omega)^2$, 
which is $O(V^2)$ as expected.

\section{Retarded Green function in perturbation theory}
\label{app:green_func_PT}
One can write explicitly the form of the retarded Green function to first order 
in perturbation theory. Proceeding in the standard way by first 
expressing the equation of motion of the evolution operator in the interaction picture and then 
expanding to first order in $V/\Omega$ we obtain 
$U_{\v{k}}(t,t') \approx e^{- i H_0(\v{k}) (t-t')}(1- i\int_{t'}^{t}~ds H_{ext}^{I}(s,t'))$
where $H_{ext}^{I}$ is the perturbing Hamiltonian in the interaction picture with respect to $H_0$.
Using the expression  $g^r_{\alpha\beta}(\v{k},t,t') = -i \theta(t-t') U_{\v{k},\alpha\beta}(t,t')$
and expanding the Green function in Pauli matrices,
$g^r=g_0^r  + g_i^r \sigma_i$ we obtain for circular polarization,

\begin{align}
g^{r,(0)}_0(k_x,\bar{t},t_r) &= - i \Theta(t_r) \cos (v k_x t_r), \nn  \\
g^{r,(1)}_0(k_x,\bar{t},t_r) &= - \frac{2 i V}{\Omega} \Theta(t_r) \sin (vk_x t_r) \sin (\frac{t_r\Omega}{2})\cos (\Omega\bar{t}), \nn
\end{align}
where we have retained only the part proportional to the identity as only this part 
contributes to the spectral function after taking the trace. We have set $k_y=0$ for simplicity to illustrate our point.
After Fourier transforming in $t_r$ we note that to zeroth order, we obtain the usual Dirac-like 
dispersion $\pm vk_x$ corresponding to the 
eigenvalues of $H_0$ and to first order the Green's function is explicitly T-periodic 
in $\bar{t}$, with sharply defined excitation bands at $\pm vk_x \pm \Omega/2$.

\bibliographystyle{apsrev}

\begin{thebibliography}{37}
\expandafter\ifx\csname natexlab\endcsname\relax\def\natexlab#1{#1}\fi
\expandafter\ifx\csname bibnamefont\endcsname\relax
  \def\bibnamefont#1{#1}\fi
\expandafter\ifx\csname bibfnamefont\endcsname\relax
  \def\bibfnamefont#1{#1}\fi
\expandafter\ifx\csname citenamefont\endcsname\relax
  \def\citenamefont#1{#1}\fi
\expandafter\ifx\csname url\endcsname\relax
  \def\url#1{\texttt{#1}}\fi
\expandafter\ifx\csname urlprefix\endcsname\relax\def\urlprefix{URL }\fi
\providecommand{\bibinfo}[2]{#2}
\providecommand{\eprint}[2][]{\url{#2}}

\bibitem[{\citenamefont{Tsui et~al.}(1982)\citenamefont{Tsui, Stormer, and
  Gossard}}]{Tsui1982}
\bibinfo{author}{\bibfnamefont{D.~C.} \bibnamefont{Tsui}},
  \bibinfo{author}{\bibfnamefont{H.~L.} \bibnamefont{Stormer}},
  \bibnamefont{and} \bibinfo{author}{\bibfnamefont{A.~C.}
  \bibnamefont{Gossard}}, \bibinfo{journal}{Phys. Rev. Lett.}
  \textbf{\bibinfo{volume}{48}}, \bibinfo{pages}{1559} (\bibinfo{year}{1982}).

\bibitem[{\citenamefont{Volovik}(1992)}]{Volovik1992}
\bibinfo{author}{\bibfnamefont{G.~E.} \bibnamefont{Volovik}},
  \emph{\bibinfo{title}{Exotic Properties of Superfluid $^3$He}}
  (\bibinfo{publisher}{World Scientific, Singapore}, \bibinfo{year}{1992}).

\bibitem[{\citenamefont{Su et~al.}(1979)\citenamefont{Su, Schrieffer, and
  Heeger}}]{Su1979}
\bibinfo{author}{\bibfnamefont{W.~P.} \bibnamefont{Su}},
  \bibinfo{author}{\bibfnamefont{J.~R.} \bibnamefont{Schrieffer}},
  \bibnamefont{and} \bibinfo{author}{\bibfnamefont{A.~J.}
  \bibnamefont{Heeger}}, \bibinfo{journal}{Phy. Rev. Lett.}
  \textbf{\bibinfo{volume}{42}}, \bibinfo{pages}{1698} (\bibinfo{year}{1979}).

\bibitem[{\citenamefont{Hasan and Kane}(2010)}]{Hasan2010}
\bibinfo{author}{\bibfnamefont{M.~Z.} \bibnamefont{Hasan}} \bibnamefont{and}
  \bibinfo{author}{\bibfnamefont{C.~L.} \bibnamefont{Kane}},
  \bibinfo{journal}{Rev. Mod. Phys.} \textbf{\bibinfo{volume}{82}},
  \bibinfo{pages}{3045} (\bibinfo{year}{2010}).

\bibitem[{\citenamefont{Roy}(2009)}]{Roy2009}
\bibinfo{author}{\bibfnamefont{R.}~\bibnamefont{Roy}}, \bibinfo{journal}{Phys.
  Rev. B} \textbf{\bibinfo{volume}{79}}, \bibinfo{pages}{195322}
  (\bibinfo{year}{2009}).

\bibitem[{\citenamefont{Moore and Balents}(2007)}]{Moore2007}
\bibinfo{author}{\bibfnamefont{J.~E.} \bibnamefont{Moore}} \bibnamefont{and}
  \bibinfo{author}{\bibfnamefont{L.}~\bibnamefont{Balents}},
  \bibinfo{journal}{Phys. Rev. B} \textbf{\bibinfo{volume}{75}},
  \bibinfo{pages}{121306} (\bibinfo{year}{2007}).

\bibitem[{\citenamefont{Fu and Kane}(2007)}]{Fu2007}
\bibinfo{author}{\bibfnamefont{L.}~\bibnamefont{Fu}} \bibnamefont{and}
  \bibinfo{author}{\bibfnamefont{C.~L.} \bibnamefont{Kane}},
  \bibinfo{journal}{Phys. Rev. B} \textbf{\bibinfo{volume}{76}},
  \bibinfo{pages}{045302} (\bibinfo{year}{2007}).

\bibitem[{\citenamefont{Dzero et~al.}(2010)\citenamefont{Dzero, Sun, Galitski,
  and Coleman}}]{Dzero2010}
\bibinfo{author}{\bibfnamefont{M.}~\bibnamefont{Dzero}},
  \bibinfo{author}{\bibfnamefont{K.}~\bibnamefont{Sun}},
  \bibinfo{author}{\bibfnamefont{V.}~\bibnamefont{Galitski}}, \bibnamefont{and}
  \bibinfo{author}{\bibfnamefont{P.}~\bibnamefont{Coleman}},
  \bibinfo{journal}{Phys. Rev. Lett.} \textbf{\bibinfo{volume}{104}},
  \bibinfo{pages}{106408} (\bibinfo{year}{2010}).

\bibitem[{\citenamefont{Wolgast et~al.}()\citenamefont{Wolgast, Kurdak, Sun,
  Allen, Kim, and Fisk}}]{Wolgast}
\bibinfo{author}{\bibfnamefont{S.}~\bibnamefont{Wolgast}},
  \bibinfo{author}{\bibfnamefont{C.}~\bibnamefont{Kurdak}},
  \bibinfo{author}{\bibfnamefont{K.}~\bibnamefont{Sun}},
  \bibinfo{author}{\bibfnamefont{J.~W.} \bibnamefont{Allen}},
  \bibinfo{author}{\bibfnamefont{D.-J.} \bibnamefont{Kim}}, \bibnamefont{and}
  \bibinfo{author}{\bibfnamefont{Z.}~\bibnamefont{Fisk}},
  \bibinfo{note}{arXiv:1211.5104 [cond-mat.str-el]}.

\bibitem[{\citenamefont{Zhang et~al.}(2013)\citenamefont{Zhang, Butch, Syers,
  Ziemak, Greene, and Paglione}}]{Zhang2013}
\bibinfo{author}{\bibfnamefont{X.}~\bibnamefont{Zhang}},
  \bibinfo{author}{\bibfnamefont{N.~P.} \bibnamefont{Butch}},
  \bibinfo{author}{\bibfnamefont{P.}~\bibnamefont{Syers}},
  \bibinfo{author}{\bibfnamefont{S.}~\bibnamefont{Ziemak}},
  \bibinfo{author}{\bibfnamefont{R.~L.} \bibnamefont{Greene}},
  \bibnamefont{and} \bibinfo{author}{\bibfnamefont{J.}~\bibnamefont{Paglione}},
  \bibinfo{journal}{Phys. Rev. X} \textbf{\bibinfo{volume}{3}},
  \bibinfo{pages}{011011} (\bibinfo{year}{2013}).

\bibitem[{\citenamefont{Lindner et~al.}(2011)\citenamefont{Lindner, Refael, and
  Galitski}}]{Lindner2011}
\bibinfo{author}{\bibfnamefont{N.~H.} \bibnamefont{Lindner}},
  \bibinfo{author}{\bibfnamefont{G.}~\bibnamefont{Refael}}, \bibnamefont{and}
  \bibinfo{author}{\bibfnamefont{V.}~\bibnamefont{Galitski}},
  \bibinfo{journal}{Nature Physics} \textbf{\bibinfo{volume}{7}},
  \bibinfo{pages}{490} (\bibinfo{year}{2011}).

\bibitem[{\citenamefont{Kitagawa et~al.}(2010)\citenamefont{Kitagawa, Berg,
  Rudner, and Demler}}]{Kitagawa2010}
\bibinfo{author}{\bibfnamefont{T.}~\bibnamefont{Kitagawa}},
  \bibinfo{author}{\bibfnamefont{E.}~\bibnamefont{Berg}},
  \bibinfo{author}{\bibfnamefont{M.}~\bibnamefont{Rudner}}, \bibnamefont{and}
  \bibinfo{author}{\bibfnamefont{E.}~\bibnamefont{Demler}},
  \bibinfo{journal}{Phys. Rev. B} \textbf{\bibinfo{volume}{82}},
  \bibinfo{pages}{235114} (\bibinfo{year}{2010}).

\bibitem[{\citenamefont{Rudner et~al.}(2013)\citenamefont{Rudner, Lindner,
  Berg, and Levin}}]{Rudner2013}
\bibinfo{author}{\bibfnamefont{M.~S.} \bibnamefont{Rudner}},
  \bibinfo{author}{\bibfnamefont{N.~H.} \bibnamefont{Lindner}},
  \bibinfo{author}{\bibfnamefont{E.}~\bibnamefont{Berg}}, \bibnamefont{and}
  \bibinfo{author}{\bibfnamefont{M.}~\bibnamefont{Levin}},
  \bibinfo{journal}{Phys. Rev. X} \textbf{\bibinfo{volume}{3}},
  \bibinfo{pages}{031005} (\bibinfo{year}{2013}).

\bibitem[{\citenamefont{Rechtsman et~al.}(2013)\citenamefont{Rechtsman, Zeuner,
  Plotnik, Lumer, Podolsky, Dreisow, Nolte, Segev, and
  Szameit}}]{Rechtsman2013}
\bibinfo{author}{\bibfnamefont{M.~C.} \bibnamefont{Rechtsman}},
  \bibinfo{author}{\bibfnamefont{J.~M.} \bibnamefont{Zeuner}},
  \bibinfo{author}{\bibfnamefont{Y.}~\bibnamefont{Plotnik}},
  \bibinfo{author}{\bibfnamefont{Y.}~\bibnamefont{Lumer}},
  \bibinfo{author}{\bibfnamefont{D.}~\bibnamefont{Podolsky}},
  \bibinfo{author}{\bibfnamefont{F.}~\bibnamefont{Dreisow}},
  \bibinfo{author}{\bibfnamefont{S.}~\bibnamefont{Nolte}},
  \bibinfo{author}{\bibfnamefont{M.}~\bibnamefont{Segev}}, \bibnamefont{and}
  \bibinfo{author}{\bibfnamefont{A.}~\bibnamefont{Szameit}},
  \bibinfo{journal}{Nature (London)} \textbf{\bibinfo{volume}{496}},
  \bibinfo{pages}{196} (\bibinfo{year}{2013}).

\bibitem[{\citenamefont{Kitagawa et~al.}(2011)\citenamefont{Kitagawa, Oka,
  Brataas, Fu, and Demler}}]{Kitagawa2011}
\bibinfo{author}{\bibfnamefont{T.}~\bibnamefont{Kitagawa}},
  \bibinfo{author}{\bibfnamefont{T.}~\bibnamefont{Oka}},
  \bibinfo{author}{\bibfnamefont{A.}~\bibnamefont{Brataas}},
  \bibinfo{author}{\bibfnamefont{L.}~\bibnamefont{Fu}}, \bibnamefont{and}
  \bibinfo{author}{\bibfnamefont{E.}~\bibnamefont{Demler}},
  \bibinfo{journal}{Phys. Rev. B} \textbf{\bibinfo{volume}{84}},
  \bibinfo{pages}{235108} (\bibinfo{year}{2011}).

\bibitem[{\citenamefont{Oka and Aoki}(2009)}]{Oka2009}
\bibinfo{author}{\bibfnamefont{T.}~\bibnamefont{Oka}} \bibnamefont{and}
  \bibinfo{author}{\bibfnamefont{H.}~\bibnamefont{Aoki}},
  \bibinfo{journal}{Phys. Rev. B} \textbf{\bibinfo{volume}{79}},
  \bibinfo{pages}{081406} (\bibinfo{year}{2009}).

\bibitem[{\citenamefont{Gu et~al.}(2011)\citenamefont{Gu, Fertig, Arovas, and
  Auerbach}}]{Gu2011}
\bibinfo{author}{\bibfnamefont{Z.}~\bibnamefont{Gu}},
  \bibinfo{author}{\bibfnamefont{H.~A.} \bibnamefont{Fertig}},
  \bibinfo{author}{\bibfnamefont{D.~P.} \bibnamefont{Arovas}},
  \bibnamefont{and} \bibinfo{author}{\bibfnamefont{A.}~\bibnamefont{Auerbach}},
  \bibinfo{journal}{Phy. Rev. Lett.} \textbf{\bibinfo{volume}{107}},
  \bibinfo{pages}{216601} (\bibinfo{year}{2011}).

\bibitem[{\citenamefont{D\'{o}ra et~al.}(2012)\citenamefont{D\'{o}ra, Cayssol,
  Simon, and Moessner}}]{Dora2012}
\bibinfo{author}{\bibfnamefont{B.}~\bibnamefont{D\'{o}ra}},
  \bibinfo{author}{\bibfnamefont{J.}~\bibnamefont{Cayssol}},
  \bibinfo{author}{\bibfnamefont{F.}~\bibnamefont{Simon}}, \bibnamefont{and}
  \bibinfo{author}{\bibfnamefont{R.}~\bibnamefont{Moessner}},
  \bibinfo{journal}{Phys. Rev. Lett.} \textbf{\bibinfo{volume}{108}},
  \bibinfo{pages}{056602} (\bibinfo{year}{2012}).

\bibitem[{\citenamefont{Wang et~al.}(2012)\citenamefont{Wang, Hsieh, Sie,
  Steinberg, Gardner, Lee, Jarillo-Herrero, and Gedik}}]{Wang2012}
\bibinfo{author}{\bibfnamefont{Y.~H.} \bibnamefont{Wang}},
  \bibinfo{author}{\bibfnamefont{D.}~\bibnamefont{Hsieh}},
  \bibinfo{author}{\bibfnamefont{E.~J.} \bibnamefont{Sie}},
  \bibinfo{author}{\bibfnamefont{H.}~\bibnamefont{Steinberg}},
  \bibinfo{author}{\bibfnamefont{D.~R.} \bibnamefont{Gardner}},
  \bibinfo{author}{\bibfnamefont{Y.~S.} \bibnamefont{Lee}},
  \bibinfo{author}{\bibfnamefont{P.}~\bibnamefont{Jarillo-Herrero}},
  \bibnamefont{and} \bibinfo{author}{\bibfnamefont{N.}~\bibnamefont{Gedik}},
  \bibinfo{journal}{Phys. Rev. Lett.} \textbf{\bibinfo{volume}{109}},
  \bibinfo{pages}{127401} (\bibinfo{year}{2012}).

\bibitem[{\citenamefont{Hsieh et~al.}(2008)\citenamefont{Hsieh, Qian, Wray,
  Xia, Hor, Cava, and Hasan}}]{Hsieh2008}
\bibinfo{author}{\bibfnamefont{D.}~\bibnamefont{Hsieh}},
  \bibinfo{author}{\bibfnamefont{D.}~\bibnamefont{Qian}},
  \bibinfo{author}{\bibfnamefont{L.}~\bibnamefont{Wray}},
  \bibinfo{author}{\bibfnamefont{Y.}~\bibnamefont{Xia}},
  \bibinfo{author}{\bibfnamefont{Y.~S.} \bibnamefont{Hor}},
  \bibinfo{author}{\bibfnamefont{R.~J.} \bibnamefont{Cava}}, \bibnamefont{and}
  \bibinfo{author}{\bibfnamefont{M.~Z.} \bibnamefont{Hasan}},
  \bibinfo{journal}{Nature (London)} \textbf{\bibinfo{volume}{452}},
  \bibinfo{pages}{970} (\bibinfo{year}{2008}).

\bibitem[{\citenamefont{Xia et~al.}(2009)\citenamefont{Xia, Qian, Hsieh, Wray,
  Pal, Lin, Bansil, Grauer, Hor, Cava et~al.}}]{Xia2009}
\bibinfo{author}{\bibfnamefont{Y.}~\bibnamefont{Xia}},
  \bibinfo{author}{\bibfnamefont{D.}~\bibnamefont{Qian}},
  \bibinfo{author}{\bibfnamefont{D.}~\bibnamefont{Hsieh}},
  \bibinfo{author}{\bibfnamefont{L.}~\bibnamefont{Wray}},
  \bibinfo{author}{\bibfnamefont{A.}~\bibnamefont{Pal}},
  \bibinfo{author}{\bibfnamefont{H.}~\bibnamefont{Lin}},
  \bibinfo{author}{\bibfnamefont{A.}~\bibnamefont{Bansil}},
  \bibinfo{author}{\bibfnamefont{D.}~\bibnamefont{Grauer}},
  \bibinfo{author}{\bibfnamefont{Y.~S.} \bibnamefont{Hor}},
  \bibinfo{author}{\bibfnamefont{R.~J.} \bibnamefont{Cava}},
  \bibnamefont{et~al.}, \bibinfo{journal}{Nat. Phys.}
  \textbf{\bibinfo{volume}{5}}, \bibinfo{pages}{398} (\bibinfo{year}{2009}).

\bibitem[{\citenamefont{Zhang et~al.}(2009)\citenamefont{Zhang, Liu, Qi, Dai,
  Fang, and Zhang}}]{Zhang2009}
\bibinfo{author}{\bibfnamefont{H.}~\bibnamefont{Zhang}},
  \bibinfo{author}{\bibfnamefont{C.-X.} \bibnamefont{Liu}},
  \bibinfo{author}{\bibfnamefont{X.-L.} \bibnamefont{Qi}},
  \bibinfo{author}{\bibfnamefont{X.}~\bibnamefont{Dai}},
  \bibinfo{author}{\bibfnamefont{Z.}~\bibnamefont{Fang}}, \bibnamefont{and}
  \bibinfo{author}{\bibfnamefont{S.-C.} \bibnamefont{Zhang}},
  \bibinfo{journal}{Nat. Phys.} \textbf{\bibinfo{volume}{5}},
  \bibinfo{pages}{438} (\bibinfo{year}{2009}).

\bibitem[{\citenamefont{Syzranov et~al.}(2008)\citenamefont{Syzranov, Fistul,
  and Efetov}}]{Syzranov2008}
\bibinfo{author}{\bibfnamefont{S.~V.} \bibnamefont{Syzranov}},
  \bibinfo{author}{\bibfnamefont{M.~V.} \bibnamefont{Fistul}},
  \bibnamefont{and} \bibinfo{author}{\bibfnamefont{K.~B.}
  \bibnamefont{Efetov}}, \bibinfo{journal}{Phys. Rev. B}
  \textbf{\bibinfo{volume}{78}}, \bibinfo{pages}{045407}
  (\bibinfo{year}{2008}).

\bibitem[{\citenamefont{Zhou and Wu}(2011)}]{Zhou2011}
\bibinfo{author}{\bibfnamefont{Y.}~\bibnamefont{Zhou}} \bibnamefont{and}
  \bibinfo{author}{\bibfnamefont{M.~W.} \bibnamefont{Wu}},
  \bibinfo{journal}{Phys. Rev. B} \textbf{\bibinfo{volume}{83}},
  \bibinfo{pages}{245436} (\bibinfo{year}{2011}).

\bibitem[{\citenamefont{G{\'o}mez-Le{\'o}n and Platero}(2013)}]{Gomez-Leon2013}
\bibinfo{author}{\bibfnamefont{A.}~\bibnamefont{G{\'o}mez-Le{\'o}n}}
  \bibnamefont{and} \bibinfo{author}{\bibfnamefont{G.}~\bibnamefont{Platero}},
  \bibinfo{journal}{Phys. Rev. Lett.} \textbf{\bibinfo{volume}{110}},
  \bibinfo{pages}{200403} (\bibinfo{year}{2013}).

\bibitem[{\citenamefont{Wang et~al.}()\citenamefont{Wang, Steinberg,
  Jarillo-Herrero, and Gedik}}]{Wang}
\bibinfo{author}{\bibfnamefont{Y.~H.} \bibnamefont{Wang}},
  \bibinfo{author}{\bibfnamefont{H.}~\bibnamefont{Steinberg}},
  \bibinfo{author}{\bibfnamefont{P.}~\bibnamefont{Jarillo-Herrero}},
  \bibnamefont{and} \bibinfo{author}{\bibfnamefont{N.}~\bibnamefont{Gedik}},
  \bibinfo{note}{submitted}.

\bibitem[{\citenamefont{Kadanoff and Baym}(1989)}]{Kadanoff1989}
\bibinfo{author}{\bibfnamefont{L.~P.} \bibnamefont{Kadanoff}} \bibnamefont{and}
  \bibinfo{author}{\bibfnamefont{G.}~\bibnamefont{Baym}},
  \emph{\bibinfo{title}{Quantum Statistical Mechanics}}
  (\bibinfo{publisher}{Perseus Books, Cambridge, Mass}, \bibinfo{year}{1989}).

\bibitem[{\citenamefont{Shirley}(1965)}]{Shirley1965}
\bibinfo{author}{\bibfnamefont{J.~H.} \bibnamefont{Shirley}},
  \bibinfo{journal}{Phys. Rev.} \textbf{\bibinfo{volume}{138}},
  \bibinfo{pages}{979} (\bibinfo{year}{1965}).

\bibitem[{\citenamefont{Turkowski and J.K.Freericks}(2007)}]{Turkowski2007}
\bibinfo{author}{\bibfnamefont{V.}~\bibnamefont{Turkowski}} \bibnamefont{and}
  \bibinfo{author}{\bibnamefont{J.K.Freericks}}, \emph{\bibinfo{title}{Strongly
  Correlated Systems, Coherence and Entanglement}} (\bibinfo{publisher}{World
  Scientific, Singapore}, \bibinfo{year}{2007}).

\bibitem[{\citenamefont{Freericks and Joura}(2008)}]{Freericks2008}
\bibinfo{author}{\bibfnamefont{J.~K.} \bibnamefont{Freericks}}
  \bibnamefont{and} \bibinfo{author}{\bibfnamefont{A.~V.} \bibnamefont{Joura}},
  \emph{\bibinfo{title}{Electron transport in nanosystems}}
  (\bibinfo{publisher}{Springer, Berlin}, \bibinfo{year}{2008}).

\bibitem[{\citenamefont{Tsuji et~al.}(2008)\citenamefont{Tsuji, Oka, and
  Aoki}}]{Tsuji2008}
\bibinfo{author}{\bibfnamefont{N.}~\bibnamefont{Tsuji}},
  \bibinfo{author}{\bibfnamefont{T.}~\bibnamefont{Oka}}, \bibnamefont{and}
  \bibinfo{author}{\bibfnamefont{H.}~\bibnamefont{Aoki}},
  \bibinfo{journal}{Phys. Rev. B} \textbf{\bibinfo{volume}{78}},
  \bibinfo{pages}{235124} (\bibinfo{year}{2008}).

\bibitem[{\citenamefont{Dittrich et~al.}(1998)\citenamefont{Dittrich,
  H{\"a}nggi, Ingold, Kramer, Shon, and Zwerger}}]{Dittrich1998}
\bibinfo{author}{\bibfnamefont{T.}~\bibnamefont{Dittrich}},
  \bibinfo{author}{\bibfnamefont{P.}~\bibnamefont{H{\"a}nggi}},
  \bibinfo{author}{\bibfnamefont{G.-L.} \bibnamefont{Ingold}},
  \bibinfo{author}{\bibfnamefont{B.}~\bibnamefont{Kramer}},
  \bibinfo{author}{\bibfnamefont{G.}~\bibnamefont{Shon}}, \bibnamefont{and}
  \bibinfo{author}{\bibfnamefont{W.}~\bibnamefont{Zwerger}},
  \emph{\bibinfo{title}{Quantum transport and dissipation}}
  (\bibinfo{publisher}{Wiley-vch, Weinheim}, \bibinfo{year}{1998}).

\bibitem[{\citenamefont{Galitski}(2011)}]{Galitski2011}
\bibinfo{author}{\bibfnamefont{V.}~\bibnamefont{Galitski}},
  \bibinfo{journal}{Phys. Rev. A} \textbf{\bibinfo{volume}{84}},
  \bibinfo{pages}{012118} (\bibinfo{year}{2011}).

\bibitem[{\citenamefont{Wilson et~al.}(2012)\citenamefont{Wilson, Fregoso, and
  Galitski}}]{Wilson2012}
\bibinfo{author}{\bibfnamefont{J.~H.} \bibnamefont{Wilson}},
  \bibinfo{author}{\bibfnamefont{B.~M.} \bibnamefont{Fregoso}},
  \bibnamefont{and} \bibinfo{author}{\bibfnamefont{V.~M.}
  \bibnamefont{Galitski}}, \bibinfo{journal}{Phys. Rev. B}
  \textbf{\bibinfo{volume}{85}}, \bibinfo{pages}{174304}
  (\bibinfo{year}{2012}).

\bibitem[{\citenamefont{Gro{\ss}mann and H{\"a}nggi}(1992)}]{Grossmann1992}
\bibinfo{author}{\bibfnamefont{F.}~\bibnamefont{Gro{\ss}mann}}
  \bibnamefont{and}
  \bibinfo{author}{\bibfnamefont{P.}~\bibnamefont{H{\"a}nggi}},
  \bibinfo{journal}{Europhys. Lett.} \textbf{\bibinfo{volume}{18}},
  \bibinfo{pages}{571} (\bibinfo{year}{1992}).

\bibitem[{\citenamefont{Freericks et~al.}(2009)\citenamefont{Freericks,
  Krishnamurty, and Pruschke}}]{Freericks2009}
\bibinfo{author}{\bibfnamefont{J.~K.} \bibnamefont{Freericks}},
  \bibinfo{author}{\bibfnamefont{H.~R.} \bibnamefont{Krishnamurty}},
  \bibnamefont{and} \bibinfo{author}{\bibfnamefont{T.}~\bibnamefont{Pruschke}},
  \bibinfo{journal}{Phys. Rev. Lett.} \textbf{\bibinfo{volume}{102}},
  \bibinfo{pages}{136401} (\bibinfo{year}{2009}).

\bibitem[{\citenamefont{Delplace et~al.}()\citenamefont{Delplace,
  G{\'o}mez-Le{\'o}n, and Platero}}]{Delplace}
\bibinfo{author}{\bibfnamefont{P.}~\bibnamefont{Delplace}},
  \bibinfo{author}{\bibfnamefont{{\'A}.}~\bibnamefont{G{\'o}mez-Le{\'o}n}},
  \bibnamefont{and} \bibinfo{author}{\bibfnamefont{G.}~\bibnamefont{Platero}},
  \bibinfo{note}{arXiv:1304.6272 [cond-mat.mes-hall]}.

\end{thebibliography}

\end{document}